\documentstyle[amssymb,aps,12pt]{revtex}

\newlength{\defaultparindent}
\begin{document}
\author{David D. Reid, Daniel W. Kittell, Eric E. Arsznov, and Gregory B. Thompson}
\address{Department of Physics and Astronomy, Eastern Michigan University, Ypsilanti,%
\\
MI 48197\vspace{0.4in}}
\title{The picture of our universe: A view from modern cosmology}
\maketitle

\begin{abstract}
In this paper we give a pedagogical review of the recent observational
results in cosmology from the study of type Ia supernovae and anisotropies
in the cosmic microwave background. By providing consistent constraints on
the cosmological parameters, these results paint a concrete picture of our
present-day universe. We present this new picture and show how it can be
used to answer some of the basic questions that cosmologists have been
asking for several decades. This paper is most appropriate for students of
general relativity and/or relativistic cosmology.
\end{abstract}

\section{INTRODUCTION}

Since the time that Einstein pioneered relativistic cosmology, the field of
cosmology has been dominated by theoretical considerations that have ranged
from straightforward applications of well-understood physics to some of the
most fanciful ideas in all of science. However, in the last several years
observational cosmology has taken the forefront. In particular, the results
of recent observations on high-redshift supernovae and anisotropies in the
radiation from the cosmic microwave background (CMB) have pinned down the
major cosmological parameters to sufficient accuracy that a precise picture
of our universe has now emerged. In this paper, we present this picture as
currently suggested by the beautiful marriage of theory and experiment that
now lies at the heart of modern cosmology.

We begin our discussion, in sections II and III, with a review of the
standard theory of the present-day universe that persisted, virtually
unaltered, from the time of Einstein until the mid 1990s. This review will
lay most of the theoretical groundwork needed for sections IV and V on the
two experimental efforts that has had such a major impact over the last few
years. Once the new results have been explained, we present, in section VI,
the picture of our universe that has emerged from the recent results. We
then conclude this paper with some brief comments on the implications of
these results for our understanding of not just the present-day universe,
but of its past and future. To help make this discussion more accessible, we
use SI units with time measured in seconds instead of meters and with all
factors of $G$ and $c$ explicitly shown unless otherwise noted.

\section{REVIEW OF THE STANDARD PRESENTATION OF COSMOLOGY}

The basic tenet that governs cosmology is known as the {\it cosmological
principle}. This principle states that, on large scales, the present
universe is homogeneous and isotropic. Homogeneity means that the properties
of the universe are the same everywhere in the universe; and isotropy means
that from every point, the properties of the universe are the same in every
direction. It can be shown that the cosmological principle alone requires
that the metric tensor of the universe must take the form of the
Robertson-Walker metric [1]. In co-moving, spherical coordinates, this
metric tensor leads to the well known line element 
\begin{equation}
ds^2=c^2dt^2-a^2(t)\left[ \frac{dr^2}{1-kr^2}+r^2d\theta ^2+r^2\sin ^2\theta
d\phi ^2\right] ,
\end{equation}
where the dimensionless function $a(t)$ is called the cosmic scale factor.
This line element describes an expanding (or contracting) universe that, at
the present (or any) instant in time $t_0$, is a three-dimensional
hypersphere of constant scalar curvature $K(t_0)=k/a^2(t_0)$. The parameter $%
k$ represents the sign of this constant which can either be positive ($k=+1$
m$^{-2}$), negative ($k=-1$ m$^{-2}$), or zero ($k=0$).

The dynamics of the universe is governed by the Einstein field equations 
\begin{equation}
R_{\mu \nu }-\frac 12g_{\mu \nu }R=\frac{8\pi G}{c^4}T_{\mu \nu },
\end{equation}
where $R_{\mu \nu }$ is the Ricci tensor, $R$ is the scalar curvature, and $%
T_{\mu \nu }$ is the stress-energy tensor. On large scales, the
stress-energy tensor of the universe is taken to be that of a perfect fluid
(since homogeneity and isotropy imply that there is no bulk energy
transport) 
\begin{equation}
T_{\mu \nu }=(p+\rho )u_\mu u_\nu /c^2-pg_{\mu \nu },
\end{equation}
where $u_\mu $ is the four-velocity of the fluid, $\rho $ is its energy
density, and $p$ is the fluid pressure. In Eqs. (2) and (3), $g_{\mu \nu }$
is relative to the Cartesian coordinates ($x^0=ct,x^1=x,x^2=y,x^3=z$).

Under the restrictions imposed by the cosmological principle, the field
equations (2) reduce to the Friedmann equations for the cosmic scale factor 
\begin{equation}
\frac{\stackrel{\cdot \cdot }{a}}a=-\frac{4\pi G}{3c^2}(\rho +3p)
\end{equation}
\begin{equation}
\left( \frac{\stackrel{\cdot }{a}}a\right) ^2=\frac{8\pi G}{3c^2}\rho -\frac{%
kc^2}{a^2},
\end{equation}
where the dot notation represents a derivative with respect to time, as is
customary. Observationally, it is known that the universe is expanding. The
expansion of the universe follows the Hubble law 
\begin{equation}
v_r=\frac{\stackrel{\cdot }{a}}ad=Hd,
\end{equation}
where $v_r$ is the speed of recession between two points, $d$ is the proper
distance between these points, and $H$ ($\equiv \stackrel{\cdot }{a}/a$) is
called the {\it Hubble parameter}. One of the key features of the Hubble law
is that, at any given instant, the speed of recession is directly
proportional to the distance. Therefore, by analyzing the Doppler shift in
the light from a distant source we can infer its distance provided that we
know the present value of the Hubble parameter $H_0$, called the {\it Hubble
constant} [2].

One of the principal questions that the field of cosmology hopes to answer
concerns the ultimate fate of the universe. Will the universe expand
forever, or will the expansion halt and be followed by a contraction? This
question of the long-term fate of the expansion is closely connected to the
sign of $k$ in the Friedmann equation (5). General relativity teaches us
that the curvature of spacetime is determined by the density of matter and
energy. Therefore, the two terms on the right-hand-side of Eq. (5) are not
independent. The value of the energy density will determine the curvature of
spacetime and, consequently, the ultimate fate of the expansion. Recognizing
that the left-hand-side of Eq. (5) is $H^2$, we can rewrite this expression
as 
\begin{equation}
1=\frac{8\pi G\rho }{3c^2H^2}-\frac{kc^2}{H^2a^2},
\end{equation}
and make the following definition: 
\begin{equation}
\Omega \equiv \frac{8\pi G\rho }{3c^2H^2},
\end{equation}
called the {\it density parameter}.

Since the sum of the density and curvature terms in Eq. (7) equals unity,
the case for which $\Omega <1$ corresponds to a negative curvature term
requiring $k=-1$ m$^{-2}$. The solution for negative curvature is such that
the universe expands forever with excess velocity $\stackrel{\cdot }{a}%
_{t=\infty }>0$. This latter case is referred to as an {\it open} universe.
The case for which $\Omega >1$ corresponds to a positive curvature term
requiring $k=+1$ m$^{-2}$. The solution for positive curvature (a {\it closed%
} universe) is such that the expansion eventually halts and becomes a
universal contraction leading to what is known as the {\it big crunch}.
Finally, the case for which $\Omega =1$ corresponds to zero curvature (a 
{\it flat} universe) requiring $k=0$. The solution for a flat universe is
the critical case that lies on the boundary between an open and closed
universe. In this case, the universe expands forever, but the rate of
expansion approaches zero asymptotically, $\stackrel{\cdot }{a}_{t=\infty
}=0 $. The value of the energy density for which $\Omega =1$ is called the 
{\it critical density} $\rho _c$, given by 
\begin{equation}
\rho _c=\frac{3c^2H^2}{8\pi G}.
\end{equation}
The density parameter, then, is the ratio of the energy density of the
universe to the critical density $\Omega =\rho /\rho _c$.

There is one final parameter that is used to characterize the universal
expansion. Notice that Eq. (4) expresses the basic result that in a
matter-dominated universe (in which $\rho +3p>0$) the expansion should be
decelerating, $\stackrel{\cdot \cdot }{a}<0$, as a result of the collective
gravitational attraction of the matter and energy in the universe. This
behavior is characterized by the {\it deceleration parameter} 
\begin{equation}
q\equiv -\frac{\stackrel{\cdot \cdot }{a}a}{\stackrel{\cdot }{a}^2}.
\end{equation}
The matter in the present universe is very sparse, so that it is effectively
noninteracting (i.e., dust). Therefore, it is generally assumed that the
fluid pressure, $p$, is negligible compared to the energy density. Under
these conditions, Eq. (4) shows that the deceleration parameter has a
straightforward relationship to the density parameter 
\begin{equation}
q=\Omega /2.
\end{equation}
Collectively, the Hubble constant $H_0$ and the present values of the
density parameter $\Omega _0$ and the deceleration parameter $q_0$ are known
as the {\it cosmological parameters}. These parameters are chosen, in part,
because they are potentially measurable. The range of values that correspond
to the different fates of the universe, within this traditional framework,
are summarized in Table 1.

\begin{center}
TABLE 1. The ranges of the main cosmological parameters for the three

\hspace{-0.45in}models of the universe in the standard presentation of
cosmology
\end{center}

\[
\begin{tabular}{p{0.75in}p{0.75in}p{0.75in}p{0.75in}}
\hline\hline
\quad Model & \multicolumn{3}{c}{Parameters} \\ \hline
& \multicolumn{3}{l}{$50<H_0<100$ km$\cdot $s$^{-1}\cdot $Mpc$^{-1}$} \\ 
\quad Open & $\Omega <1$ & $\rho _m<\rho _c$ & $q_0<1/2$ \\ 
\quad Closed & $\Omega >1$ & $\rho _m>\rho _c$ & $q_0>1/2$ \\ 
\quad Flat & $\Omega =1$ & $\rho _m=\rho _c$ & $q_0=1/2$ \\ \hline\hline
\end{tabular}
\]
\medskip

The understanding of cosmology, as outlined above, left several questions
unanswered, including the fate of the universal expansion. It was entirely
possible to formulate reasonable arguments for whether or not the universe
is open, closed, or flat that covered all three possibilities. The
observational data has always suggested that the density of visible matter
is insufficient to close the universe and researchers choosing to side with
the data could easily take the position that the universe is open. However,
it has been known for several decades that a substantial amount of the
matter in the universe, perhaps even most of it, is not visible. The
existence of large amounts of {\it dark matter} can be inferred from its
gravitational effects both on and within galaxies [3]. Therefore, the
prospects of dark matter (and neutrino mass) rendered any conclusion based
solely on the amount of visible matter premature. Einstein's view was that
the universe is closed, apparently for reasons having to do with Mach's
principle [4], and many researchers preferred this view as well for reasons
that were sometimes more philosophical than scientific. Then, there were
also hints that the universe may be flat; consequentially, many researchers
believed that this was most likely true.

Belief that the universe is flat was partly justified by what is known in
cosmology as the {\it flatness problem} having to do with the apparent need
of the universe to have been exceedingly close to the critical density
shortly after the big bang. The fate of the universe and the flatness
problem are just two of several puzzles that emerge from this standard model
of the universe. Another important puzzle has to do with the existence, or
nonexistence, of the {\it cosmological constant}, $\Lambda $. It turns out
that $\Lambda $ plays a very important role in our story, and {\it its}
story must be told before we can explain how cosmologists have pinned down
some of the basic properties of the universe.

\section{THE COSMOLOGICAL CONSTANT}

In 1915 Albert Einstein introduced his theory of General Relativity. Like
Newton before him, Einstein's desire was to apply his theory to cosmology.
Einstein embraced the prevailing view at that time that the universe is
static. Therefore, he attempted to find solutions of the form $\stackrel{%
\cdot }{a}=0$. It soon became apparent that even with Einstein's theory of
gravity, as with Newton's, the gravitational attraction of the matter in the
universe causes a static universe to be unstable. Furthermore, as can be
seen from Eq. (4), the subsequent requirement of $\stackrel{\cdot \cdot }{a}%
=0$ implies a negative pressure such that $p=-\rho /3$. For ordinary stellar
matter and gas, this relationship is not physically reasonable.

To remedy such problems, Einstein modified his original field equations from
Eq. (2) to the more general form 
\begin{equation}
R_{\mu \nu }-\frac 12g_{\mu \nu }R-\Lambda g_{\mu \nu }=\frac{8\pi G}{c^4}%
T_{\mu \nu },
\end{equation}
where $\Lambda $ is the cosmological constant mentioned in the previous
section. Equation (12) is the most general form of the field equations that
remains consistent with the physical requirements of a relativistic theory
of gravity. The cosmological constant term, for $\Lambda >0$, can be viewed
as a repulsive form of gravity that is independent of the curvature of
spacetime. The modern approach is to treat $\Lambda $ as a form of energy
present even in empty space - vacuum energy [5]. This interpretation implies
modifying Eq. (12) to 
\begin{equation}
R_{\mu \nu }-\frac 12g_{\mu \nu }R=\frac{8\pi G}{c^4}\left( T_{\mu \nu }+%
\frac{\Lambda c^4}{8\pi G}g_{\mu \nu }\right) .
\end{equation}
In the perfect fluid approximation, this leads to an effective fluid
pressure and energy density given by 
\begin{equation}
p=p_m-\frac{\Lambda c^4}{8\pi G}
\end{equation}
\begin{equation}
\rho =\rho _m+\frac{\Lambda c^4}{8\pi G},
\end{equation}
where $p_m$ and $\rho _m$ are the pressure and energy density of the {\it %
matter} content of the universe. As Eq. (14) shows, the cosmological
constant contributes a negative term to the pressure in the universe. This
affect of the cosmological constant allowed Einstein to find a static,
albeit unstable, solution for the dynamics of the universe.

Once it became known that the universe is expanding, Einstein discarded the
cosmological constant term having no other physical reason to include it.
However, the possible existence of a non-zero cosmological constant has been
a subject of debate ever since. With the cosmological constant in the
picture, the equations for the dynamics of the universe, Eqs. (4) and (5),
generalize to 
\begin{equation}
\frac{\stackrel{\cdot \cdot }{a}}a=-\frac{4\pi G}{3c^2}(\rho _m+3p_m)+\frac{%
\Lambda c^2}3
\end{equation}
\begin{equation}
\left( \frac{\stackrel{\cdot }{a}}a\right) ^2=H^2=\frac{8\pi G}{3c^2}\rho _m-%
\frac{kc^2}{a^2}+\frac{\Lambda c^2}3.
\end{equation}

Besides Einstein's static model of the universe, another interesting, and
important, solution to Eqs. (16) and (17), known as the {\it de Sitter
solution}, applies to the case of a spatially flat, empty universe ($\rho
_m=0,$ $p_m=0,$ $k=0$). In this case, $\Lambda $ supplies the only
contribution to the energy density 
\begin{equation}
\rho =-p=\frac{\Lambda c^4}{8\pi G}.
\end{equation}
Equation (16) shows that under these conditions the universe would be
accelerating $\stackrel{..}{a}>0$, and Eq. (17) shows that the Hubble
parameter would be given by 
\begin{equation}
H=\left( \frac{\Lambda c^2}3\right) ^{1/2}.
\end{equation}
The de Sitter solution for the cosmic scale factor shows that the effect of
the cosmological constant is to cause the accelerating universe to expand
exponentially with time according to 
\begin{equation}
a(t)=a_0e^{Ht}.
\end{equation}

Obviously, the universe is not completely empty, but the de Sitter solution
remains important because it is possible for a cosmological constant term to
be sufficiently large as to dominate the dynamics of the universe. The
dominant components of the universe are determined by the relative values of
the corresponding density parameters. Dividing Eq. (17) by $H^2$ produces
the analog of Eq. (7) 
\begin{equation}
1=\frac{8\pi G\rho _m}{3H^2c^2}+\frac{kc^2}{a^2H^2}+\frac{\Lambda c^2}{3H^2},
\end{equation}
where the sign of $k$ has be separated out. The first term on the
right-hand-side is called the matter term and, in analogy with Eq. (8), also
gives the matter density parameter $\Omega _m$. The second term is the
curvature term and is characterized by the curvature density parameter $%
\Omega _k$. Finally, the last term is known as the vacuum-energy density
parameter $\Omega _\Lambda $. Thus, in a universe with a cosmological
constant, the primary density parameters are 
\begin{equation}
\Omega _m=\frac{8\pi G\rho _m}{3H^2c^2},\hspace{0.2in}\Omega _k=-\frac{kc^2}{%
a^2H^2},\hspace{0.2in}\Omega _\Lambda =\frac{\Lambda c^2}{3H^2}.
\end{equation}
The present values of these parameters, together with the Hubble constant,
would determine the dynamics of the universe in this model [6].

\section{DETERMINING COSMOLOGICAL PARAMETERS FROM TYPE IA SUPERNOVAE}

\subsection{Type Ia Supernovae}

Throughout their lives, stars remain in stable (hydrostatic) equilibrium due
to the balance between outward pressures (from the fluid and radiation) and
the inward pressure due to the gravitational force. The enormously energetic
nuclear fusion that occurs in stellar cores causes the outward pressure. The
weight of the outer region of the star causes the inward pressure. A
supernova occurs when the gravitational pressure overcomes the internal
pressure, causing the star to collapse, and then violently explode. There is
so much energy released (in the form of light) that we can see these events
out to extremely large distances.

Supernovae are classified into two types according to their spectral
features and light curves (plot of luminosity vs. time). Specifically, the
spectra of type Ia supernovae are hydrogen-poor, and their light curves show
a sharp rise with a steady, gradual decline. In addition to these
spectroscopic features, the locations of these supernovae, and the absence
of planetary nebulae, allow us to determine the genesis of these events.
Based on these facts, it is believed that the progenitor of a type Ia
supernova is a binary star system consisting of a {\it white dwarf} with a 
{\it red giant} companion [7]. Other binary systems have been theorized to
cause these supernovae, but are not consistent with spectroscopic
observation [8].

Although the Sun is not part of a binary system, approximately half of all
stellar systems are. Both members are gravitationally bound and therefore
revolve around each other. While a binary star system is very common, the
members of the progenitor to a type Ia supernova have special properties.
White dwarf stars are different from stars like the Sun in that nuclear
fusion does not take place within these objects. Electron degeneracy
pressure, which is related to the well known Pauli exclusion principle,
holds the white dwarf up against its own weight. For electron degeneracy
pressure to become important, an object must be extremely dense. White dwarf
stars have the mass of the Sun, but are the size of the Earth. Also, the
physics of this exotic form of pressure produces a strange effect: heavier
white dwarfs are actually smaller in size (mass $\times $ volume = constant)
[9]. Red giant stars, on the other hand, are the largest known stars and
contain a relatively small amount of mass. As a result, gravity is
relatively weak at the exterior region of red giant stars.

In such a binary system, the strong gravitational attraction of the white
dwarf overcomes the weaker gravity of the red giant. At the outer edge of
the red giant, the gravitational force from the white dwarf is stronger than
that from the red giant. This causes mass from the outer envelope of the red
giant to be accreted onto the white dwarf. As a result, the mass of the
white dwarf increases, causing its size to decrease. This process continues
until the mass of the white dwarf reaches the {\it Chandrasekhar limit}
(1.44 solar masses) beyond which electron degeneracy pressure is no longer
able to balance the increasing pressure due to the gravitational force. At
the center of the white dwarf, the intense pressure and temperature ignites
the fusion of Carbon nuclei. This sudden burst of energy produces an
explosive deflagration (subsonic) wave that destroys the star. This
violently exploding white dwarf is what we see as a type Ia supernova.

The use of type Ia supernovae for determining cosmological parameters rests
on the ability of these supernovae to act as standard candles. Standard
candles have been used to determine distances to celestial objects for many
years. They are luminous objects whose intrinsic (or absolute) brightness
can be determined independent of their distance. The intrinsic brightness,
together with the observed apparent brightness (which depends on the
distance to the object), can be used to calculate distances. The distance
calculated from measurements of the luminosity (power output) of an object
is appropriately termed the {\it luminosity distance} 
\begin{equation}
d=10^{(m-M-25)/5},
\end{equation}
where $m$ is the apparent brightness measured in magnitudes (apparent
magnitude), $M$ is the absolute magnitude, and $d$ is the luminosity
distance in units of megaparsecs. The quantity, $m-M$ is commonly known as
the {\it distance modulus}. For the reader who is unfamiliar with the
magnitude scale see chapter 3 of Ref. 9.

As explained above, all type Ia supernovae are caused by the same process, a
white dwarf reaching 1.44 solar masses by accretion from a red giant. As a
result of this consistency, we not only expect to see extremely consistent
light curves from these events, but we also expect that these light curves
will reach the same peak magnitude. If this latter point is true, type Ia
supernovae can be used as standard candles and, therefore, distance
indicators.

Methods for determining the absolute magnitude of a type Ia supernova can be
divided into two categories depending on whether or not we know the distance
to the event. If we know the distance to the host galaxy of the supernova,
by means of a Cepheid variable for example, and we observe the apparent
magnitude of the event $m$, then we can use the distance modulus to
calculate the absolute magnitude directly 
\begin{equation}
m-M=5\log (d)+25.
\end{equation}
If the distance is not known, the peak luminosity must be inferred from
observational data. The techniques for making this inference often involve
corrections for many processes that would otherwise adversely affect the
results. These processes include interstellar extinction within the host
galaxy, redshift of the light from the expansion of the universe,
gravitational lensing, and an apparently natural scatter in the peak
brightness; see Ref. 10 for a discussion of these corrections. Once the
luminosity $L$ of a supernova has been determined, this luminosity, together
with the luminosity $L^{\prime }$, and absolute magnitude $M^{\prime }$, of
a well-known object (such as the Sun) will yield the absolute magnitude of
the supernova 
\begin{equation}
M=M^{\prime }-2.5\log (L/L^{\prime }).
\end{equation}
Taking all of this into account, it has been determined that the peak
absolute magnitude of type Ia supernovae is [11] 
\begin{equation}
M_{Ia}=-19.5\pm 0.2\text{ mag.}
\end{equation}

\subsection{Measuring the Hubble Constant}

As stated previously, the expansion of the universe follows the Hubble law
given by Eq. (6). Observationally, we measure the recession velocity as a
redshift, $z$, in the light from the supernova ($v_r=cz$). Since every type
Ia supernovae has about the same absolute magnitude, Eq. (26), the apparent
magnitude provides an indirect measure of its distance. Therefore, for
nearby supernovae ($z\leq 0.3$) the Hubble Law is equivalent to a
relationship between the redshift and the magnitude. Inserting (26) into
(24), using (6), and applying to the current epoch, yields the {\it %
redshift-magnitude relation} 
\begin{equation}
m=M_{Ia}+5\log (cz)-5\log (H_0)+25.
\end{equation}
Defining the $z=0$ intercept as 
\begin{equation}
\stackrel{\sim }{M}\equiv M_{Ia}-5\log (H_0)+25,
\end{equation}
we can write equation (27) as 
\begin{equation}
m=\stackrel{\sim }{M}+5\log (cz).
\end{equation}
As shown in Fig. 1, low-redshift data can be used to find $\stackrel{\sim }{M%
}$ and Eq. (28) to solve for the Hubble constant. Studies on type Ia
supernova [12] consistently suggest a value for the Hubble constant of about
63 km$\cdot $s$^{-1}\cdot $Mpc$^{-1}$.

The result for $H_0$, found from low-redshift supernovae, tends to set the
lower bound when compared with other methods for obtaining $H_0$. For
example, if the distances to enough galaxies can be accurately found, then
the Hubble law can be used directly to obtain a value of $H_0$. This has
partly been the goal of the {\it Hubble Space Telescope Key Project} [13].
This project has shown that a careful consideration of the type Ia supernova
results in combination with the other methods for obtaining $H_0$ produces
what has become a widely accepted value for the Hubble constant 
\begin{equation}
H_0=72\pm 8\text{ km}\cdot \text{s}^{-1}\cdot \text{Mpc}^{-1}.
\end{equation}
The value given in Eq. (30) is the one that we shall adopt in this paper.

\subsection{Measuring $\Omega _m$, $\Omega _\Lambda $, and $q_0$}

In order to determine the other cosmological parameters from the supernova
data we must consider supernova at large distances ($z\geqslant 0.3$). Just
as large distance measurements on Earth show us the curvature (geometry) of
Earth's surface, so do large distance measurements in cosmology show us the
geometry of the universe. Since, as we have seen, the geometry of the
universe depends on the values of the cosmological parameters, measurements
of the luminosity distance for distant supernova can be used to extract
these values.

To obtain the general expression for the luminosity distance, consider
photons from a distant source moving radially toward us. Since we are
considering photons, $ds^2=0$, and since they are moving radially, $d\theta
^2=d\phi ^2=0$. The Robertson-Walker metric, Eq. (1), then reduces to $%
0=c^2dt^2-a^2dr^2(1-kr^2)^{-1}$, which implies 
\begin{equation}
dt=\frac{adr}{c(1-kr^2)^{1/2}}.
\end{equation}

To get another expression for $dt$, we multiply Eq. (17) by $a^2(t)$ which
produces an expression for $(da/dt)^2$. Furthermore, we note that since the
universe is expanding, the matter density is a function of time. Given that
lengths scale as $a(t)$, volumes scale as $a^3(t)$ and therefore, 
\begin{equation}
\rho _m(t)\propto 1/a^3(t).
\end{equation}
Using these facts, together with the definitions of the density parameters
in Eq. (22), Eq. (17) becomes 
\begin{equation}
\left( \frac{da}{dt}\right) ^2=H_0^2\left[ \Omega _{m,0}\frac{a_0}a+\Omega
_{k,0}+\Omega _{\Lambda ,0}\left( \frac a{a_0}\right) ^2\right] .
\end{equation}
As previously mentioned, it is better to write things in terms of measurable
quantities, and in this case we can directly relate the cosmic scale factor
to the redshift $z$. The redshift is defined such that 
\begin{equation}
1+z=\frac{\lambda _0}\lambda ,
\end{equation}
where $\lambda _0$ is the current (received) value of the wavelength and $%
\lambda $ is the wavelength at the time of emission. The redshift is a
direct result of the cosmic expansion and it can be shown that [14] $\lambda
\propto a(t)$; therefore, 
\begin{equation}
\frac{a_0}a=1+z.
\end{equation}
Using Eq. (35) and the fact that $\Omega _k=1-\Omega _m-\Omega _\Lambda $
from Eq. (21), Eq. (33) can be rewritten as 
\begin{equation}
dt=H_0^{-1}(1+z)^{-1}\left[ (1+z)^2(1+\Omega _{m,0}z)-z(z+2)\Omega _{\Lambda
,0}\right] ^{-1/2}dz.
\end{equation}
Equating the expressions in Eqs. (31) and (36) and integrating, leads to an
expression for the radial coordinate $r$ of the star. The luminosity
distance is then given by [15] $d=(1+z)a_0r$. Therefore, 
\begin{equation}
d=\frac{c(1+z)}{H_0|\Omega _{k,0}|^{1/2}}\text{sinn}\left\{ |\Omega
_{k,0}|^{1/2}\int\nolimits_0^z\left[ (1+z^{\prime })^2(1+\Omega
_{m,0}z^{\prime })-z^{\prime }(z^{\prime }+2)\Omega _{\Lambda ,0}\right]
^{-1/2}dz^{\prime }\right\} ,
\end{equation}
where sinn$(x)$ is $\sinh (x)$ for $k<0$, $\sin (x)$ for $k>0$, and if $k=0$
neither sinn nor $|\Omega _{k,0}|$ appear in the expression. We see that the
functional dependence of the luminosity distance is $d(z;\Omega _m,\Omega
_\Lambda )$.

Inserting Eq. (37) into Eq. (24), and using the intercept from Eq. (28), we
get a redshift-magnitude relation valid at high z 
\begin{equation}
m-\stackrel{\sim }{M}=5\log [d(z;\Omega _m,\Omega _\Lambda )]
\end{equation}
In practice, astronomers observe the apparent magnitude and redshift of a
distant supernova. The density parameters are then determined by those
values that produce the best fit to the observed data according to Eq. (38)
for different cosmological models.

Under the continued assumption that the fluid pressure of the matter in the
universe is negligible ($p_m\approx 0$), Eq. (16) implies that the
deceleration parameter at the present time is given by 
\begin{equation}
q_0=\Omega _{m,0}/2-\Omega _{\Lambda ,0}.
\end{equation}
Therefore, once the density parameters have been determined by the above
procedure, the deceleration parameter can then be found.

Figure 2 illustrates how high-redshift data can be used to estimate the
cosmological parameters and provide evidence in favor of a nonzero
cosmological constant. In this figure, the abscissa is the difference
between the distance moduli for the observed supernovae and what would be
expected for a traditional cosmological model such as those represented in
Table 1. The case shown is based on the data of Riess {\it et. al.} [16]
using a traditional model with $\Omega _m=0.2$ and $\Omega _\Lambda =0$
represented by the central line $\Delta (m-M)=0$. The figure shows that the
data points lie predominantly above the zero line. This result means that
the supernovae are further away (or equivalently, dimmer) than traditional,
decelerating cosmological models allow. The conclusion then is that the
universe must be accelerating. As suggested by Eq. (39), the most
straightforward explanation of this conclusion is the presence of a nonzero,
positive cosmological constant. The solid curve, above the zero line in Fig.
2, represents a best-fit curve to the data that corresponds to a universe
with $\Omega _m=0.24$ and $\Omega _\Lambda =0.72$.

Typical values for the cosmological parameters as determined by detailed
analysis of the type just discussed are the following [16]: 
\begin{eqnarray}
&&\Omega _{m,0}=0.24_{-0.24}^{+0.56}  \nonumber \\
&&\Omega _{\Lambda ,0}=0.72_{-0.48}^{+0.72} \\
&&q_0=-1.0\pm 0.4.  \nonumber
\end{eqnarray}
Note that the negative deceleration parameter is consistent with an
accelerating universe. Furthermore, these values imply that the universe is
effectively flat predicting a curvature parameter roughly centered around $%
\Omega _k\approx 0.04$.

\section{DETERMINING COSMOLOGICAL PARAMETERS FROM ANISOTROPIES IN THE CMB}

The theory of the anisotropies in the CMB is rich with details about the
contents and structure of the early universe. Consequently, this theory can
become quite complicated. However, because of this same richness, this
branch of cosmology holds the potential to provide meaningful constraints on
a very large number of quantities of cosmological interest. Our focus here
is to provide the reader with a conceptual understanding of why and how CMB
anisotriopies can be used to determine cosmological parameters. We will
place particular emphasis on the density parameters corresponding to the
spatial curvature of the universe $\Omega _k$, and the baryon density $%
\Omega _b$. The reader seeking more detail should consult Ref. 17 and the
references therein.

\subsection{Anisotropies in the CMB}

The ``hot big bang'' model is widely accepted as the standard model of the
early universe. According to this idea, our universe started in a very hot,
very dense state that suddenly began to expand, and the expansion is
continuing today. All of space was contained in that dense point. It is not
possible to observe the expansion from an outside vantagepoint and it is not
correct to think of the big bang as happening at one point in space. The big
bang happened everywhere at once.

During the first fraction of a second after the big bang, it is widely
believed that the universe went through a brief phase of exponential
expansion called {\it inflation }[18]. Baryonic matter formed in about the
first second; and the nuclei of the light elements began to form
(nucleosynthesis) when the universe was only several minutes old. Baryons
are particles made up of three quarks; the most familiar baryons are the
protons and neutrons in the nuclei of atoms. Since all of the matter that we
normally encounter is made up of atoms, baryonic matter is considered to be
the ``ordinary'' matter in the universe.

The very early universe was hot enough to keep matter ionized, so the
universe was filled with nucleons and free electrons. The density of free
electrons was so high that Thomson scattering effectively made the universe
opaque to electromagnetic radiation. The universe remained a baryonic plasma
until around 300,000 years after the big bang when the universe had expanded
and cooled to approximately 3000 K. At this point, the universe was
sufficiently cool that the free electrons could join with protons to form
neutral hydrogen. This process is called {\it recombination}. With electrons
being taken up by atoms, the density of free electrons became sufficiently
low that the mean free path of the photons became much larger (on the order
of the size of the universe); and light was free to propagate. The light
that was freed during recombination has now cooled to a temperature of about 
$T_o=2.73$ K. This light is what we observe today as the cosmic microwave
background. We see the CMB as if it were coming from a spherical shell
called the {\it surface of last scattering} (Fig. 3). This shell has a
finite thickness because recombination occurred over a finite amount of time.

Today, over very large scales, the universe is homogeneous. However, as
evidenced by our own existence, and the existence of galaxies and groups of
galaxies, etc., inhomogeneities exist up to scales on the order of 100 Mpc.
Theories of structure formation require that the seeds of the structure we
observe today must have been inhomogeneities in the matter density of the
early universe. These inhomogeneities would have left their imprint in the
CMB which we would observe today as temperature anisotropies. So, in order
to explain the universe in which we live, there should be bumps in the CMB;
and these bumps should occur over angular scales that correspond to the
scale of observed structure. In 1992, the COBE satellite measured
temperature fluctuations $\delta T$ in the CMB, $\delta T/T\sim 10^{-5}$ on
a 7$^{\circ }$ angular scale [19], where $T$ is the ambient temperature of
the CMB. The anisotropies detected by COBE are considered to be large-scale
variations caused by nonuniformities generated at the creation of the
universe. However, recent observations [20-22] have found small-scale
anisotropies that correspond to the physical scale of today's observed
structure. It is believed that these latter anisotropies are the result of
quantum fluctuations in density that existed prior to inflation which were
greatly amplified during inflation. These amplified fluctuations became the
intrinsic density perturbations which are the seeds of structure formation.

The small-scale anisotropies in the CMB can be separated into two
categories: primary and secondary. Primary anisotropies are due to effects
that occur at the time of recombination and are ``imprinted'' in the CMB as
the photons leave the surface of last scattering. Secondary anisotropies
arise through scattering along the line of sight between the surface of last
scattering and the observer. In this paper, we will only be concerned with
the primary anisotropies. There are three main sources for primary
anisotropies in the microwave background. These are the Sachs-Wolfe effect,
intrinsic (adiabatic) perturbations, and a Doppler effect.

For the largest of these primary anisotropies the dominant mechanism is the
Sachs-Wolfe effect. At the surface of last scattering, matter density
fluctuations will lead to perturbations in the gravitational potential, $%
\delta \Phi $. These perturbations cause a gravitational redshift of the
photons coming from the surface of last scattering as they ``climb out'' of
the potential wells. This effect is described by, $\delta T/T=\delta \Phi
/c^2$. These same perturbations in the gravitational potential also cause a
time dilation at the surface of last scattering, so these photons appear to
come from a younger, hotter universe. This effect is described by, $\delta
T/T=-2(\delta \Phi )/3c^2$. Combining these two processes gives the
Sachs-Wolfe effect [23], 
\begin{equation}
\frac{\delta T}T=\frac{\delta \Phi }{3c^2}.
\end{equation}

On intermediate scales, the main effect is due to adiabatic perturbations.
Recombination occurs later in regions of higher density, so photons
emanating from overly dense regions experience a smaller redshift from the
universal expansion and thus appear hotter. The observed temperature
anisotropy resulting from this process is given by [23], 
\begin{equation}
\left( \frac{\delta T}T\right) _{obs}=-\frac{\delta z}{1+z}=\frac{\delta
\rho }\rho .
\end{equation}

Finally, on smaller scales there is a Doppler effect that becomes important.
This effect arises because the photons are last scattered in a moving
plasma. The temperature anisotropy corresponding to this effect is described
by [23], 
\begin{equation}
\frac{\delta T}T=\frac{\delta \overrightarrow{v}\cdot \widehat{r}}c,
\end{equation}
where $\widehat{r}$ denotes the direction along the line of sight and $%
\overrightarrow{v}$ is a characteristic velocity of the material in the
scattering medium.

\subsection{Acoustic Peaks and the Cosmological Parameters}

The early universe was a plasma of photons and baryons and can be treated as
a single fluid [24]. Baryons fell into the gravitational potential wells
created by the density fluctuations and were compressed. This compression
gave rise to a hotter plasma thus increasing the outward radiation pressure
from the photons. Eventually, this radiation pressure halted the compression
and caused the plasma to expand (rarefy) and cool producing less radiation
pressure. With a decreased radiation pressure, the region reached the point
where gravity again dominated and produced another compression phase. Thus,
the struggle between gravity and radiation pressure set up longitudinal
(acoustic) oscillations in the photon-baryon fluid. When matter and
radiation decoupled at recombination the pattern of acoustic oscillations
became frozen into the CMB. Today, we detect the evidence of the sound waves
(regions of higher and lower density) via the primary CMB anisotropies.

It is well known that any sound wave, no matter how complicated, can be
decomposed into a superposition of wave modes of different wavenumbers $k$,
each $k$ being inversely proportional to the physical size of the
corresponding wave (its wavelength), $k\propto 1/\lambda $. Observationally,
what is seen is a projection of the sound waves onto the sky. So, the
wavelength of a particular mode $\lambda $ is observed to subtend a
particular angle $\theta $ on the sky. Therefore, to facilitate comparison
between theory and observation, instead of a Fourier decomposition of the
acoustic oscillations in terms of sines and cosines, we use an angular
decomposition (multipole expansion) in terms of Legendre polynomials $P_\ell
(\cos \theta )$. The order of the polynomial $\ell $ (related to the
multipole moments) plays a similar role for the angular decomposition as the
wavenumber $k$ does for the Fourier decomposition. For $\ell \geq 2$ the
Legendre polynomials on the interval [-1,1] are oscillating functions
containing a greater number of oscillations as $\ell $ increases. Therefore,
the value of $\ell $ is inversely proportional to the characteristic angular
size of the wave mode it describes 
\begin{equation}
\ell \propto 1/\theta .
\end{equation}

Experimentally, temperature fluctuations can be analyzed in pairs, in
directions $\widehat{n}$ and $\widehat{n^{\prime }}$ that are separated by
an angle $\theta $ so that $\widehat{n}\cdot \widehat{n^{\prime }}=\cos
\theta $. By averaging over all such pairs, under the assumption that the
fluctuations are Gaussian, we obtain the two-point correlation function, $%
C(\theta )$, which is written in terms of the multipole expansion 
\begin{equation}
\left\langle \delta T(\widehat{n})\cdot \delta T(\widehat{n^{\prime }}%
)\right\rangle \equiv C(\theta )=\sum\limits_\ell \frac{(2\ell +1)}{4\pi }%
C_\ell P_\ell (\cos \theta ),
\end{equation}
the $C_\ell $ coefficients are called the multipole moments.

As predicted, analysis of the temperature fluctuations does in fact reveal
patterns corresponding to a harmonic series of longitudinal oscillations.
The various modes correspond to the number of oscillations completed before
recombination. The longest wavelength mode, subtending the largest angular
size for the primary anisotropies, is the fundamental mode -- this was the
first mode detected. There is now strong evidence that both the 2$^{\text{nd}%
}$ and 3$^{\text{rd}}$ modes have also been observed [20-22].

The distance sound waves could have traveled in the time before
recombination is called the {\it sound horizon}, $r_s$. The sound horizon is
a fixed physical scale at the surface of last scattering. The size of the
sound horizon depends on the values of the cosmological parameters. The
distance to the surface of last scattering, $d_{sls}$, also depends on
cosmological parameters. Together, they determine the angular size of the
sound horizon (see Fig. 3) 
\begin{equation}
\theta _s\approx \frac{r_s}{d_{sls}},
\end{equation}
in the same way that the angle subtended by the planet Jupiter depends on
both its size and distance from us. Analysis of the temperature anisotropies
in the CMB determine $\theta _s$ and the cosmological parameters can be
varied in $r_s$ and $d_{sls}$ to determine the best-fit results.

We can estimate the sound horizon by the distance that sound can travel from
the big bang, $t=0$, to recombination $t_{*}$%
\begin{equation}
r_s(z_{*};\Omega _b,\Omega _r)\approx \int_0^{t_{*}}c_sdt,
\end{equation}
where $z_{*}$ is the redshift parameter at recombination ($z_{*}\approx 1100$%
) [25], $\Omega _r$ is the density parameter for radiation (photons), $c_s$
is the speed of sound in the photon-baryon fluid, given by [26] 
\begin{equation}
c_s\approx c\left[ 3\left( 1+3\Omega _b/4\Omega _r\right) \right] ^{-1/2},
\end{equation}
which depends on the baryon-to-photon density ratio, and $dt$ is determined
by an expression similar to Eq. (36), except at an epoch in which radiation
plays a more important role. The energy density of radiation scales as $\rho
_r\propto a^{-4}$ [27], so with the addition of radiation, Eq. (33)
generalizes to 
\begin{equation}
\left( \frac{da}{dt}\right) ^2=H_0^2\left[ \Omega _{r,0}\left( \frac{a_0}a%
\right) ^2+\Omega _{m,0}\frac{a_0}a+\Omega _{k,0}+\Omega _{\Lambda ,0}\left( 
\frac a{a_0}\right) ^2\right] ,
\end{equation}
which, upon using Eq. (35) and $\Omega _r+\Omega _m+\Omega _\Lambda +\Omega
_k=1$, leads to 
\begin{equation}
dt=H_0^{-1}(1+z)^{-1}\left\{ (1+z)^2(1+\Omega _{m,0}z)+z(z+2)\left[
(1+z)^2\Omega _{r,0}-\Omega _{\Lambda ,0}\right] \right\} ^{-1/2}dz.
\end{equation}

The distance to the surface of last scattering, corresponding to its angular
size, is given by what is called the angular diameter distance. It has a
simple relationship to the luminosity distance $d$ [15] given in Eq. (37) 
\begin{equation}
d_{sls}=\frac{d(z_{*};\Omega _m,\Omega _\Lambda )}{(1+z_{*})^2}.
\end{equation}
The location of the first acoustic peak is given by $\ell \approx
d_{sls}/r_s $ and is most sensitive to the curvature of the universe $\Omega
_k$.

To get a feeling for this result, we can consider a very simplified,
heuristic calculation. We will consider a prediction for the first acoustic
peak for the case of a flat universe. To leading order, the speed of sound
in the photon-baryon fluid, Eq. (48), is constant $c_s=c/\sqrt{3}$. We
further make the simplifying assumption that the early universe was
matter-dominated (there is good reason to believe that it was which will be
discussed in the next section). With these assumptions, Eqs. (49) and (47)
yield (dropping the `0' from the density parameters) 
\begin{equation}
r_s=\frac{c_s}{H_0\sqrt{\Omega _m}}\int_{z_{*}}^\infty (1+z)^{-5/2}dz,
\end{equation}
which gives 
\begin{equation}
r_s=\frac{2c_s}{3H_0\sqrt{\Omega _m}}(1+z_{*})^{-3/2}.
\end{equation}

The distance to the surface of last scattering, in our flat universe model,
will depend on both $\Omega _m$ and $\Omega _\Lambda $. Following a
procedure similar to that which lead to Eq. (37), the radial coordinate of
the surface of last scattering, $r_{sls}$ (not to be confused with $r_s$),
is determined by 
\begin{equation}
r_{sls}=\frac c{H_0}\int_0^{z_{*}}\left[ \Omega _m(1+z)^3+\Omega _\Lambda
\right] ^{-1/2}dz,
\end{equation}
which does not yield a simple result. Using a binomial expansion, the
integrand can be approximated as $\Omega _m^{-1/2}(1+z)^{-3/2}-(\Omega
_\Lambda /2\Omega _m^{3/2})(1+z)^{-9/2}$ and the integral is more easily
handled. The distance is then determined by $d_{sls}=r_{sls}/(1+z_{*})$
which gives 
\begin{equation}
d_{sls}=\frac{2c}{7H_0(1+z_{*})}\left\{ 7\Omega _m^{-1/2}-2\Omega _\Lambda
\Omega _m^{-3/2}+O[(1+z_{*})^{-1/2}]\right\} .
\end{equation}
Using $\Omega _\Lambda =1-\Omega _m$ and neglecting the higher order terms
gives 
\begin{equation}
d_{sls}\approx \frac{2c\Omega _m^{-1/2}}{7H_0(1+z_{*})}\left\{ 9-2\Omega
_m^3\right\} .
\end{equation}

Combining Eqs. (53) and (56) to get our prediction for the first acoustic
peak gives 
\begin{equation}
\ell \approx \frac{d_{sls}}{r_s}\approx 0.74\sqrt{(1+z_{*})}\left\{
9-2\Omega _m^3\right\} \approx 221.
\end{equation}
This result is consistent with the more detailed result that [28] 
\begin{equation}
\ell \approx 200/\sqrt{1-\Omega _k},
\end{equation}
where, in our calculation $\Omega _k=0$. Equation (58) suggests that a
measurement of $\ell \approx 200$ implies a flat universe. The BOOMERanG
[22] collaboration found $\ell \approx 197\pm 6$, and the MAXIMA-1 [21]
collaboration measured $\ell \approx 220$. Additional simplified
illustrations for how the cosmological parameters can be obtained from the
acoustic peak can be found in Ref. 29.

Experimental results, such as those quoted above, are determined by plotting
the power spectrum (power per logarithmic interval), $(\delta T_\ell )^2$,
given by 
\begin{equation}
(\delta T_\ell )^2=\frac{\ell (\ell +1)}{2\pi }C_\ell ,
\end{equation}
or by the square root of this quantity. The power spectrum may be quickly
calculated for a given cosmological model using a code such as CMBFAST which
is freely available online [30]. The solid curve in Fig. 4 was calculated
using CMBFAST and the data points are only a representative few included to
show the kind of agreement between theory and experiment that exists.

While the location of the first acoustic peak helps to fix $\Omega _k$,
other features of the power spectrum help to determine the baryon density.
Since baryons are the primary cause of the gravitational potential wells
that help generate the acoustic oscillations, they affect the power spectrum
in several ways. The relative heights of the peaks are an indication of $%
\Omega _b$ in that an increase in baryon density results in an enhancement
of the odd peaks. An increase in baryon density also leads to enhanced
damping at higher multipoles [31].

It is important to recognize that the constraints on cosmological parameters
obtained through this sort of analysis are correlated so that the range of
possible values of $\Omega _\Lambda $, for example, depends on what is
assumed for the possible range of values of the Hubble constant. Therefore,
it is customary to incorporate results from other observational (or
theoretical) work in the analysis of the CMB data. With this in mind, we use
the value of the Hubble constant stated in Eq. (30). Given this assumption,
a combined study of the CMB anisotropy data from the BOOMERanG [22],
MAXIMA-1 [21], and COBE-DMR [32] collaborations suggests the following
values for the two cosmological parameters being considered here [33]: 
\begin{eqnarray}
\Omega _{k,0} &=&0.11\pm 0.07  \nonumber \\
\Omega _{b,0} &=&0.062\pm 0.01.
\end{eqnarray}
As with the type Ia supernova results, the best-fit CMB results predict an
essentially flat universe. In fact, it is quite possible to adopt a model
with $\Omega _k\equiv 0$ and still obtain a very good fit to the data along
with reasonable values for the other cosmological parameters [33]. Again,
the CMB data also provides values for additional cosmological parameters,
but the curvature and baryon densities are perhaps the most accurately
constrained at this time.

Even though the recent revolution in cosmology was ignited by the type Ia
supernova and CMB anisotropy results, it is also important to acknowledge
prior work toward constraining the cosmological parameters. This work
includes investigations on gravitational lensing [34], large-scale structure
[35], and the ages of stars, galaxies, and globular clusters [36]. Without
this work, the ability to use the supernova and CMB data to place fairly
tight restrictions on the major cosmological parameters would be
significantly diminished.

\section{THE PICTURE OF OUR UNIVERSE}

Given the results from observational cosmology discussed in the previous two
sections we are now able to present a concrete picture of the universe, as
opposed to the traditional array of models with very different properties.
Taking a more comprehensive view, in Table 2 we present a set of
cosmological parameters (without errors) that might be taken as the ``best
estimates'' based on various observational and theoretical studies [37].

\begin{center}
TABLE 2. Our best estimates of the cosmological parameters

for the present-day universe and the primary sources we used

to obtain them. If ``theory'' is listed as the source, we derived

the value from other estimates by using the stated equation.\bigskip

\begin{tabular}{ccc}
\hline\hline
Parameter & Value & Primary Sources \\ \hline
Hubble Constant & $H_0=72$ km$\cdot $s$^{-1}\cdot $Mpc$^{-1}$ & [13] \\ 
Cosmological Constant & $\Omega _\Lambda =0.70$ & [16, 33] \\ 
Matter & $\Omega _m=0.30$ & [16, 33] \\ 
Baryonic matter & $\Omega _b=0.04$ & [33] \\ 
Dark matter & $\Omega _{CDM}=0.26$ & theory: Eq. (61) \\ 
Curvature & $\Omega _k=0.00$ & [16, 20-22, 33] \\ 
Deceleration parameter & $q_0=-0.55$ & theory: Eq. (39) \\ \hline\hline
\end{tabular}
\vspace{0.18in}
\end{center}

This set of parameters describes a flat universe the dynamics of which is
dominated by two mysterious forms of energy, most prominently, the
cosmological constant. So then, the long-standing debate over whether or not
the cosmological term should be included in Einstein's theory is over; not
only should it be included, it dominates the universe. Although the debate
over the existence of the cosmological constant has ended, the debate over
its physical implications has just begun. Further comments about this debate
will be discussed in the conclusion.

The other mysterious form of energy listed in Table 2, $\Omega _{CDM}$, is
dark matter where ``CDM'' stands for ``cold dark matter.'' Recall that
ordinary matter made up of atomic nuclei only contributes to the baryon
content of the universe with $\Omega _b\approx 0.04$. However, since the
total matter content is $\Omega _m\approx 0.30$, the rest of the matter in
the universe must be in some exotic, unseen form which is why we call it
dark matter 
\begin{equation}
\Omega _{CDM}=\Omega _m-\Omega _b
\end{equation}
We have known about dark matter for several decades now, having been first
discovered through anomalous rotation curves of galaxies [3]. The results
from the CMB anisotropies only help to confirm that not only does dark
matter exist, but that it comprises roughly 90\% of the matter in the
universe.

Given values of the cosmological parameters, we can now solve for the
dynamics of the universe. The Friedmann equations (16) and (17) for our
present ($p_m=0,$ $k=0$) universe can be combined to give 
\begin{equation}
2\frac{\stackrel{\cdot \cdot }{a}}a+\left( \frac{\stackrel{\cdot }{a}}a%
\right) ^2=\Lambda c^2.
\end{equation}
This equation can be solved exactly giving the result [38] 
\begin{equation}
a(t)=A^{1/3}\sinh ^{2/3}\left( \frac t{t_\Lambda }\right) ,
\end{equation}
where $A=\Omega _{m,0}/\Omega _{\Lambda ,0}\approx 0.43$ and $t_\Lambda
=(4/3\Lambda c^2)^{1/2}\approx 3.4\times 10^{17}$ s. The cosmic scale factor
is plotted in Fig. 5 and compared to the purely de Sitter universe described
by Eq. (20). From this comparison, we see that, today, the qualitative
behavior of our universe is that of a de Sitter universe except that the
presence of matter has caused the universe to expand less than in the de
Sitter case.

With $a(t)$ in hand, we can now write a precise metric for the universe 
\begin{equation}
ds^2=c^2dt^2-A^{2/3}\sinh ^{4/3}(t/t_\Lambda )\left[ dr^2+r^2d\theta
^2+r^2\sin ^2\theta d\phi ^2\right] .
\end{equation}
This tells us that we can visualize the universe as an expanding Euclidean
sphere with the expansion governed by $a(t)$ as given in Eq. (63). Note,
however, that in this visualization the universe is represented as the
entire volume of the sphere and not just the surface.

Another interesting feature that emerges from this picture is that if $%
\Lambda $ is truly constant, the universe would have once been
matter-dominated. To see why this is, recall that because the size of the
universe changes, the density parameters are functions of time. As we go
back in time, the universe gets smaller so that the energy density of matter 
$\rho _m$ gets larger while the energy density associated with $\Lambda $,
see Eq. (15), remains constant. Using Eq. (22) we can see that the ratio of
matter-to-cosmological constant is 
\begin{equation}
\frac{\Omega _m}{\Omega _\Lambda }=\frac{\Omega _{m,0}}{\Omega _{\Lambda ,0}}%
a^{-3}(t).
\end{equation}
Therefore, at some finite time in the past the universe was such that $%
\Omega _m/\Omega _\Lambda >1$. Since the expansion of a matter-dominated
universe would be decelerating, this implies that the universe underwent a
transition from decelerated expansion to accelerated expansion. This
behavior is reflected in the deceleration parameter as a function of time,
which, given the current cosmological parameters becomes 
\begin{equation}
q(t)=\frac 12\left[ 1-3\tanh ^2(t/t_\Lambda )\right] .
\end{equation}
Figure 6 is a plot of $q(t)$ and shows that the deceleration parameter was
once positive and that a transition to $q(t)<0$ occurred around the time at
which $\Omega _m/\Omega _\Lambda =1$.

Having a specific model of the universe allows us to determine specific
answers to questions that cosmologists have been asking for decades. While
we cannot address all such questions in this paper we will tackle a few of
the most common: (a) What is the age of the universe? (b) Will the universe
expand forever or will the expansion eventually stop followed by a
re-collapse? (c) Where is the edge of the observable universe?

The age of the universe can be calculated by integrating $dt$ from now, $z=0$%
, back to the beginning $z=\infty $. For our universe, the steps leading to
Eq. (36) produces 
\begin{equation}
dt=H_0^{-1}(1+z)^{-1}\left[ \Omega _{m,0}(1+z)^3+\Omega _{\Lambda ,0}\right]
^{-1/2}dz.
\end{equation}
Making the definition $x\equiv 1+z$, the present age of the universe is
given by 
\begin{equation}
t_0=H_0^{-1}\int_1^\infty \left[ \Omega _{m,0}x^5+\Omega _{\Lambda
,0}x^2\right] ^{-1/2}dx.
\end{equation}
The solution to Eq. (68) is complex. Taking only the real part gives 
\begin{equation}
t_0=\frac 2{3H_0\Omega _{\Lambda ,0}^{1/2}}\tanh ^{-1}\left[ \left( 1+\frac{%
\Omega _{m,0}}{\Omega _{\Lambda ,0}}\right) ^{1/2}\right] =13.1\times 10^9%
\text{ yr.}
\end{equation}

The question of whether or not the universe will expand forever is
determined by the asymptotic behavior of $a(t)$. Since $\sinh (x)$ diverges
as $x\rightarrow \infty $, it is clear that the universe will continue to
expand indefinitely unless some presently unknown physical process
drastically alters its dynamics.

Finally, concerning the question of the size of the observable universe,
there are two types of horizons that might fit this description, the {\it %
particle horizon} and the {\it event horizon}. The particle horizon is the
position of the most distant event that can presently be seen, that is, from
which light has had enough time to reach us since the beginning of the
universe. Unfortunately, since current evidence suggests that the universe
was not always dominated by the cosmological constant, we cannot extend the
current model back to the beginning. We can, however, extend it into the
future. The event horizon is the position of the most distant event that we
will ever see. If we consider a photon moving radially toward us from this
event, then Eq. (31) describes its flight. Since we are interested in those
events that will occur from now $t_0$, onward, Eq. (31) leads to 
\begin{equation}
\int_0^{r_{EH}}dr=cA^{-1/3}\int_{t_0}^\infty \sinh ^{-2/3}(t/t_\Lambda )dt,
\end{equation}
where $r_{EH}$ is the radial coordinate of our event horizon. Performing a
numerical solution to the integral yields 
\begin{equation}
r_{EH}\approx 1.2ct_\Lambda =16\times 10^9\text{ light years.}
\end{equation}
This result suggests that 16 billion light years is the furthest that we
will ever be able to see. As far as we are aware, the most distant object
ever observed (besides the CMB) is currently the galaxy RD1 at a redshift of 
$z=5.34$, which places it approximately 12.2 billion light years away [39].

\section{CONCLUSIONS}

In summary, the resent observational results in cosmology strongly suggest
that we live in a universe that is spatially flat, expanding at an
accelerated rate, homogeneous and isotropic on large scales, and is
approximately 13 billion years old. The expansion of the universe is
described by Eq. (63), and its metric by Eq. (64). We have seen that roughly
96\% of the matter and energy in the universe consists of cold dark matter
and the cosmological constant. We now know basic facts about the universe
much more precisely than we ever have. However, since we cannot speak with
confidence about the nature of dark matter or the cosmological constant,
perhaps the most interesting thing about all of this is that knowing more
about the universe has only shown us just how little we really understand.

As mentioned previously, the most common view of the cosmological constant
is that it is a form of vacuum energy due, perhaps, to quantum fluctuations
in spacetime [5]. However, within the context of general relativity alone
there is no need for such an interpretation; $\Lambda $ is just a natural
part of the geometric theory [40]. If, however, we adopt the view that the
cosmological constant belongs more with the energy-momentum tensor than with
the curvature tensor, this opens up a host of possibilities including the
possibility that $\Lambda $ is a function of time [41].

In conclusion, it is also important to state that although this paper
emphasizes what the recent results say about our present universe, these
results also have strong implications for our understanding of the distant
past and future of the universe. For an entertaining discussion of the
future of the universe see Ref. 42. Concerning the past, the results on
anisotropies in the CMB have provided strong evidence in favor of the
inflationary scenario, which requires a $\Lambda $-like field in the early
universe to drive the inflationary dynamics. To quote White and Cohn, ``Of
dozens of theories proposed before 1990, only inflation and cosmological
defects survived after the COBE announcement, and only inflation is
currently regarded as viable by the majority of cosmologists'' [17].

\section{ACKNOWLEDGEMENTS}

We would like to acknowledge (and recommend) the excellent website of Dr.
Wayne Hu [43]. This resource is very useful for learning about the physics
of CMB anisotropies. We are also grateful to Dr. Manasse Mbonye for making
several useful suggestions.


\begin{references}
\bibitem{2}  It should be noted that what we have called the Hubble
parameter is also often called the Hubble constant. This name is
inappropriate, however, because the Hubble parameter is time dependent.
\end{references}
\end{document}